%% file: template.tex
\documentclass[10pt,journal,cspaper,compsoc]{IEEEtran}
\ifCLASSOPTIONcompsoc
  \usepackage[nocompress]{cite}
\else
  \usepackage{cite}
\fi
\ifCLASSINFOpdf
   \usepackage[pdftex]{graphicx}
   \graphicspath{{pictures/}}
   \DeclareGraphicsExtensions{.pdf,.jpeg,.jpg,.png}
\else
\fi

\usepackage{hyperref}

\usepackage{subcaption}

\begin{document}
%
\title{Vis-a-Vis: Visual Exploration of\\Visualization Source Code Evolution}
%
%
%
%

\author{Fabian~Bolte,~\IEEEmembership{Member,~IEEE}
        and~Stefan~Bruckner,~\IEEEmembership{Member,~IEEE~Computer~Society}
\IEEEcompsocitemizethanks{\IEEEcompsocthanksitem Fabian~Bolte and Stefan~Bruckner are with the Department of Informatics, University of Bergen, Norway.\protect\\
E-mail: \{stefan.bruckner, fabian.bolte\}@uib.no
}
}

%
%

\markboth{Journal of \LaTeX\ Class Files,~Vol.~14, No.~8, August~2015}%
{Shell \MakeLowercase{\textit{et al.}}: Bare Demo of IEEEtran.cls for Computer Society Journals}
%



\IEEEtitleabstractindextext{%
\begin{abstract}
Developing an algorithm for a visualization prototype often involves the direct comparison of different development stages and design decisions, and even minor modifications may dramatically affect the results. While existing development tools provide visualizations for gaining general insight into performance and structural aspects of the source code, they neglect the central importance of result images unique to graphical algorithms. In this paper, we present a novel approach that enables visualization programmers to simultaneously explore the evolution of their algorithm during the development phase together with its corresponding visual outcomes by providing an automatically updating meta visualization. Our interactive system allows for the direct comparison of all development states on both the visual and the source code level, by providing easy to use navigation and comparison tools. The on-the-fly construction of difference images, source code differences, and a visual representation of the source code structure further enhance the user's insight into the states' interconnected changes over time. Our solution is accessible via a web-based interface that provides GPU-accelerated live execution of C++ and GLSL code, as well as supporting a domain-specific programming language for scientific visualization.%
\end{abstract}

\begin{IEEEkeywords}
Visualization System and Toolkit Design, User Interfaces, Integrating Spatial and Non-Spatial Data Visualization, Software Visualization.
\end{IEEEkeywords}
}

\maketitle

\IEEEdisplaynontitleabstractindextext

%
\IEEEpeerreviewmaketitle

\IEEEraisesectionheading{\section{Introduction}\label{sec:introduction}}

%
%
%
%
\IEEEPARstart{T}{he}
\input{content.tex}

\ifCLASSOPTIONcompsoc
\section*{Acknowledgments}
\else
\section*{Acknowledgment}
\fi

The research presented in this paper was supported by the MetaVis
project (\#250133) funded by the Research Council of Norway.

\ifCLASSOPTIONcaptionsoff
  \newpage
\fi



\bibliographystyle{IEEEtran}
\bibliography{IEEEabrv,bibliography}
%



%

\vspace{-2.75cm}
\begin{IEEEbiography}[{\includegraphics[width=1in,height=1.25in,clip,keepaspectratio]{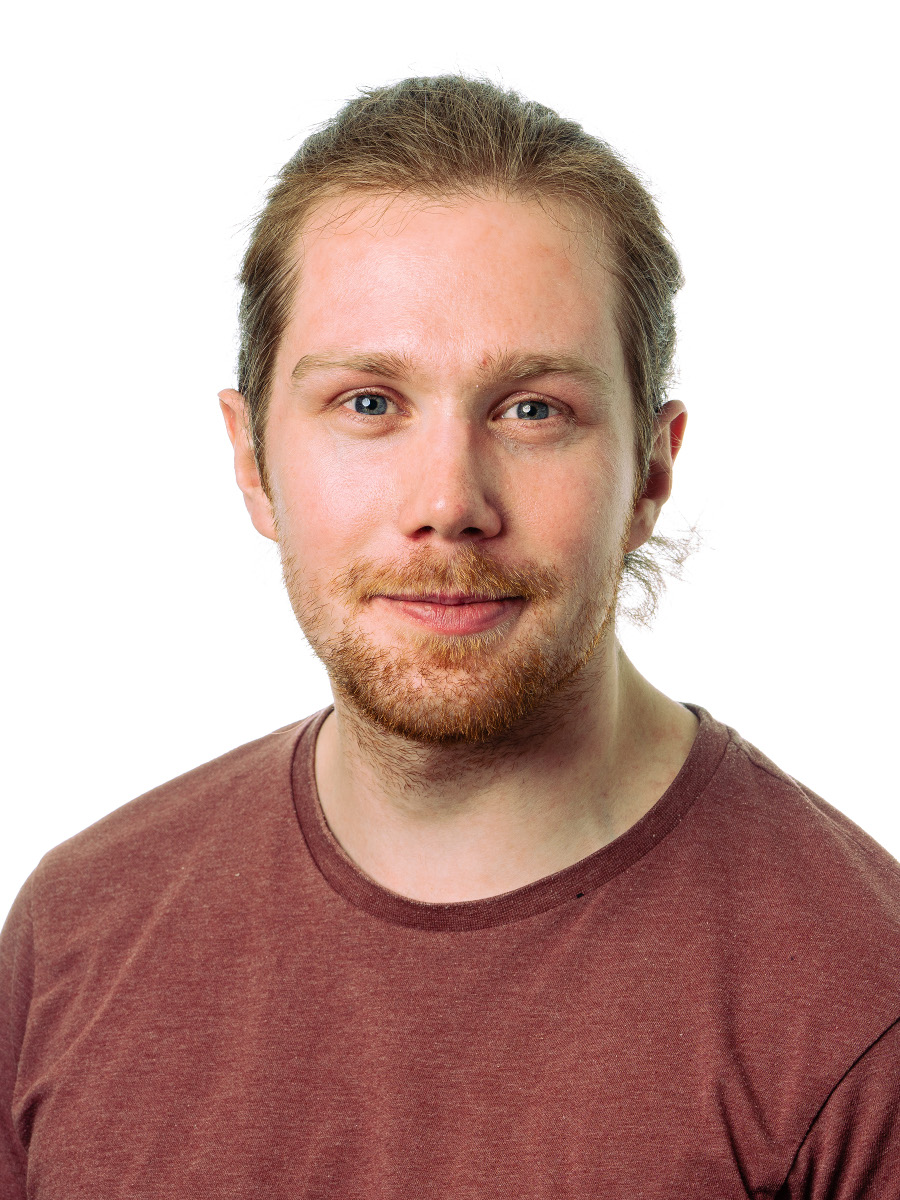}}]{Fabian Bolte}
is a PhD student in visualization at the Department of Informatics of the University of Bergen, Norway. He received his bachelor's degree in Applied Computer Science from the TU Chemnitz, Germany in 2014 and his master's degree in High Performance \& Cloud Computing in 2016 from the same university. His research interests include the visualization of time-dependent data, visual parameter space analysis, and meta visualization.
\end{IEEEbiography}

\vspace{-3cm}
\begin{IEEEbiography}[{\includegraphics[width=1in,height=1.25in,clip,keepaspectratio]{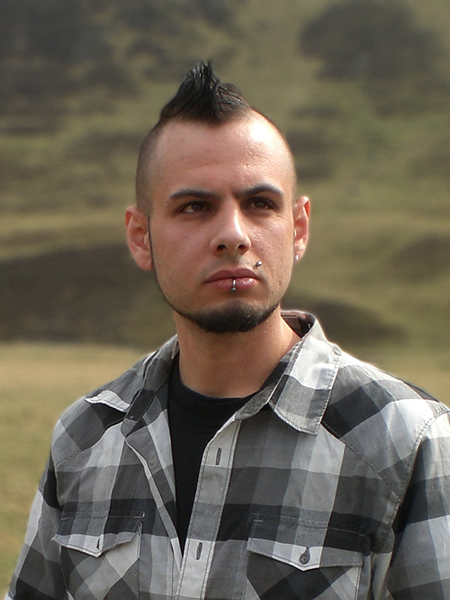}}]{Stefan Bruckner}
is professor in visualization at the Department of Informatics of the University of Bergen, Norway. He received his master’s degree in Computer Science from the TU Wien, Austria in 2004 and his Ph.D. in 2008 from the same university. He was awarded the habilitation (venia docendi) in Practical Computer Science in 2012.
From 2008 to 2013, he was an assistant professor at the Institute of Computer Graphics and Algorithms at TU Wien. His research interests include interactive visualization techniques for spatial data, particularly in the context of biomedical applications, visual parameter space analysis, illustrative methods, volume visualization, and knowledge-assisted visual interfaces.
\end{IEEEbiography}







\end{document}

%% file: content.tex
 process of developing a visualization algorithm typically involves a trial-and-error approach consisting of the repetitive task sequence of writing code, compiling the program, and comparing the visual result to previous outputs. This is common practice in research and industry for fixing bugs or developing new features. Existing development tools ease the software development on a general level by providing visual insight into the source code structure and the application's performance. Development tools specifically designed for visualization algorithms could further improve the user's experience by providing insight into the visual changes created by the source code~\cite{laramee2009using}. An overview of the visual results created at different points in time can ease the comparison of features in the algorithm. Source code changes and changes in the visual result can be investigated as a unit, instead of being considered individually. When teaching visualization algorithms to students, free exploration of the source code's evolution and its resulting visual outcome can build a deeper understanding of the underlying technical details and problems that occur during the development of such algorithms.
Some recent tools~\cite{Satyanarayan2016,ShaderToy} provide a live view of the application's result and thereby relieve the user from the common compile-and-run cycle. While several approaches present a visual history of the source code evolution~\cite{Voinea2005,Telea2008,Wittenhagen2016}, none of them connect the source code to its graphical output. The development of visualization techniques, in particular for prototyping and education purposes, could greatly benefit from a coupling of the source code to its visual result, making it easier to pinpoint when artifacts are introduced or whether a specific method is sensitive to noise in the data.

In this paper, we present a novel approach for visualizing an algorithm's evolution for a general purpose (C++, GLSL) and a domain-specific programming language targeted at scientific visualization algorithms. We provide tools for investigating all revisions at different levels of detail with side-by-side comparisons for visual results, a visual representation of the source code's structure, difference images, and source code differences. We further apply user-defined algorithmic parameters to all states of the evolution to ease the comparison task with respect to parameter changes. We present results in an automatically updating and interactive environment, that enables direct state comparison and navigation throughout the whole development process with instant state switching for free investigation. Our system provides visual support of the development process in an interactive visual analysis tool for visualization researchers and practitioners. The tool can further be utilized in the education domain to teach visualization algorithms and their detailed differences to students, building a fundamental understanding of the correlations between algorithmic and visual changes, and a strong basis for future visualization research.

Our main contributions can be summarized as follows:
\begin{itemize}
  \item We introduce a novel approach for the concurrent live visualization of the evolution of scientific visualization source code and its visual output.
  \item We provide automatic revision management with interactive state switching and visual guidance.
  \item Algorithm parameters can be automatically mapped to user interface elements and their effects can be explored interactively on the entire revision history.
  \item The system is implemented as a web-based client-server environment enabling the development of GPU-based visualization algorithms on any client.
\end{itemize}

\section{Related Work}
\label{sec:related}

Our system provides support for the development of scientific visualization prototypes and benefits from on-the-fly compilation, live previews, automatic revision management, and parameter management. While our general approach of combining these features into a single development environment is novel, many visualization techniques exist for individual features.\\

\noindent\textbf{Terminology}

\noindent
A visualization developer (user), writes \emph{visualization source code} to describe an algorithm. Instead of considering full-fledged visualization solutions, which integrate the whole visualization pipeline (reading and processing data, creating a render window, etc.) we mainly focus on the visual mapping and rendering part. We use the term \emph{state} to refer to a revision of the source code in a version control system. Every state creates a \emph{visual result} (\emph{output}) based on a given parameter set. We create a \emph{meta visualization}, used in the same sense as Bertini et al.\cite{Bertini2011} -- a visualization of visualizations, to showcase the evolution of an algorithm.\\

\noindent\textbf{Software Visualization}

\noindent
Software visualization has been a prominent topic over many years and many powerful methods have been developed. They visualize software projects at different granularities, such as the source code level, the software structure, or the runtime behavior of the program~\cite{Ball1996}.

SeeSoft~\cite{Eick1992} represents a file as a column and each line of code (LOC) as a fixed size row. Marcus et al.~\cite{Marcus2003} reduce the required space of this visualization by representing each LOC as a square and appending them to fixed length rows. They further highlight the syntactical structure of the source code by coloring each square depending on its nesting level or keywords of the language.
The visualization of source code history is likewise a prominent research field in software visualization~\cite{Merino2016a}. DeVis~\cite{Zhi2013} summarizes the number of LOCs that have been added, deleted, or modified in a pie chart and provides this information over time in a spiral layout. CVSscan~\cite{Voinea2005} highlights structures in source files and aligns the LOCs across time to create a history view. Telea and Auber~\cite{Telea2008} take this approach further by directly visualizing the evolution via a flow graph. Holten and van Wijk\cite{Holten2008} show how this approach can be enhanced by edge bundling. Chronicler~\cite{Wittenhagen2016} builds a node-link diagram from the source code structure and connects correlating nodes in a flow graph to display the evolution of hierarchical structures.
The entire structure of a software project can be visualized to gain an overview of all existing classes, their relationships~\cite{DAmbros2009} and their evolution~\cite{Collberg2003,Lanza2001,Wettel2008,Khan2011}. Additionally, the authors of developed classes and their collaborations can be displayed~\cite{German2004,Ogawa2010}.
The runtime behavior and performance of an application can be visualized to detect issues or bottlenecks within the code~\cite{Isaacs2014}.

This kind of visual analysis can be generally applied to, but is not specifically designed for, source code of visualization algorithms. In our approach, we take advantage of the visual output that is specific to visualization algorithms and provide tools to support the intrinsic needs of visualization developers.\\

\noindent\textbf{Visualization Pipeline}

\noindent
Existing frameworks like VTK~\cite{schroeder2004visualization}, ParaView~\cite{ahrens2005paraview} and MeVisLab~\cite{koenig2006embedding} enable the construction of and insight into the pipeline of a scientific visualization process. Each pipeline consists of several modules featuring algorithms with their respective inputs and outputs, including operations from data space, through visualization space, into image space. These kinds of frameworks provide visual support for the rapid prototyping and analysis of visualization pipelines. They simplify the task of comparing individual pipelines against each other and finding proper parameter sets to run them with. In contrast, our system provides visual support for the process of prototyping a visualization algorithm, which could then be utilized as a module within one of these frameworks.\\

\noindent\textbf{Parameter Management}

\noindent
Many systems provide support for finding a suitable parameter set for a given visualization. Design Galleries~\cite{marks1997design} generate multiple output images from dispersed parameter sets to provide an overview of the parameter space and enable the easy identification of desirable results.
Image graphs~\cite{ma1999image} display the parameter changes between different output images and spreadsheet-like interfaces~\cite{jankun2001visualization} allow for the investigation of parameter effects and their interplay.
Sedlmair et al.~\cite{sedlmair2014visual} evaluate existing work on visual parameter space analysis and divide problems, strategies, and tasks into a conceptual framework.

All these approaches analyze and compare the effect of different parameter sets for a given visualization with the goal of finding one suitable set. In contrast, we utilize one parameter set at a time to analyze its effect on different visualizations in order to inform the user about aspects such as sensitivity to parameter changes during different phases of development.\\

\noindent\textbf{Visual Provenance}

\noindent
Visualizing the evolution of a workflow is closely related to the notions of provenance~\cite{herschel2017survey} and graphical histories~\cite{Heer2008}. They provide important information for improving the user's understanding, the product's quality, and ensuring reproducibility. Several tools, including Kepler~\cite{Altintas}, Triana~\cite{Taylor2005}, and work by Pimentel et al.~\cite{pimentel2016tracking}, focus on the history of user interactions and modifications in data, meta data, information systems, and workflow. Such systems have shown that a version tree and thumbnails can be utilized to improve a user's exploration capabilities~\cite{stitz2018knowledgepearls,camisetty2018enhancing}.
VisTrails~\cite{Bavoil2005} is a system specifically designed for visualizing provenance in the visualization pipeline, allowing for the comparison of several visualizations for different data and parameter sets. It therefore addresses many of the issues found in the process of constructing a visualization pipeline for a given task and data set, that we face in the context of developing a visualization algorithm.
However, our approach starts at an earlier level, even before the visualization methods are known, and visualizes the provenance of the algorithms themselves by allowing for the investigation of the development process. We thereby provide a support system for visualization developers instead of users of existing techniques.\\
\\

\noindent\textbf{Live Execution \& Notebooks}

\noindent
An increasing number of powerful online tools provide live previews for the output of code in different languages.
Notebooks like Jupyter~\cite{kluyver2016jupyter} and Observable~\cite{Observable} allow for the definition and individual execution of code snippets, interlaced with text to provide explanations of the code. This leads to a tighter coupling of source code and its output for interpreted languages.
For compiled languages, on-the-fly previews, as seen in Overleaf~\cite{Overleaf} for \LaTeX, are often only provided for the output of the complete source code. ShaderToy~\cite{ShaderToy} allows for the editing of GLSL code and combines a live view of the result and predefined inputs to create a powerful environment for prototyping shader effects. Vega-Lite~\cite{Satyanarayan2016} is a high-level grammar for interactive information visualization that provides an online editor with instant visual feedback and extensions for visual debugging techniques~\cite{hoffswell2016visual}. Literate visualization~\cite{wood2019design} builds an evolution of visualizations and explanations for their design choices on top of that.
None of these solutions, however, provide means for exploring the evolution of the code and its results over time.\\

\section{Overview}
\label{sec:overview}

Visualization has been proven to be invaluable for many applications in vastly varying fields, by improving the user's insight into complex data~\cite{hansen2011visualization}. It is therefore remarkable that we, as visualization researchers and practitioners, make limited use of advanced visualization tools in what is often a considerable part of our daily work -- the development of visualization algorithms. We want to improve the current situation by providing an interactive tool for visualization developers to support the investigation of their work. Our first approach in this direction is aimed at the prototyping of such algorithms and we see great potential for our application being utilized in the education domain for teaching visualization algorithms to students.

\subsection{User and Task Requirements}

\begin{figure}[tb]
 \centering
 \includegraphics[width=\columnwidth]{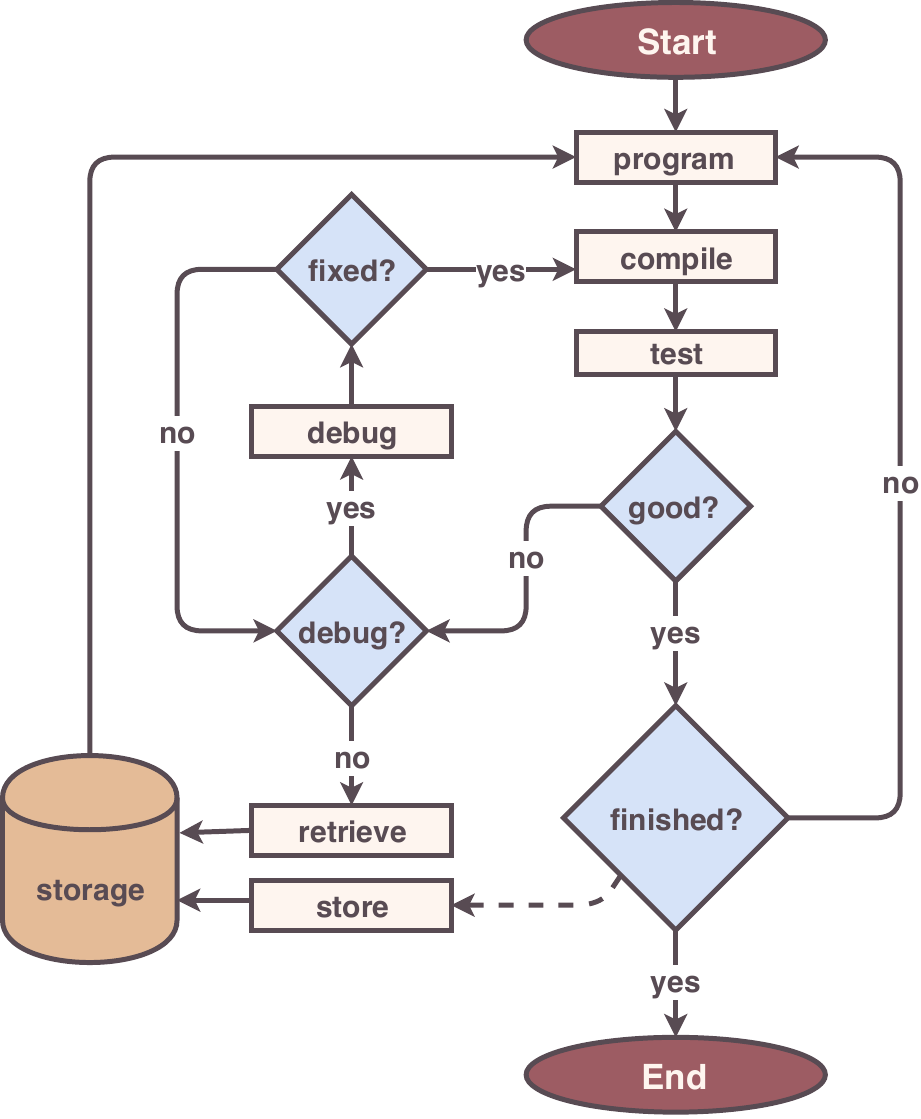}
 \caption{Development workflow. During the development process, a user programs a visualization algorithm, compiles the source code, and tests the outcome. Bugs are typically fixed in an iterative manner. The whole process is normally performed separately for each feature of the implementation. Functioning features can be manually stored, and restored when the current state is not satisfactory.}
 \label{fig:workflow}
\end{figure}

In order to improve the visualization researcher's development process, it is important to first understand their problems and needs. We therefore take a look at a programmer's standard workflow, visualized in \autoref{fig:workflow}. 
During the development process a user typically (partially) implements a visualization algorithm, compiles the source code, and tests if the developed program generates the expected visual result. If it does, the current state can be manually stored as a backup and extended until it supports all required features. If an intermediate result is incorrect or undesired in any way, the source code must be inspected to locate and fix the issue (debug), recompiled, and tested again. This time-consuming process must be repeated until the result appears free of issues. If the identified issue cannot be located or cannot be fixed, a previously stored state must be restored. Even if the source code is free of issues, the implemented feature might turn out to be unsuitable for the given task and should be replaced by another feature. This again requires the user to continue from a previous state.

Many visualization algorithms define several parameters which significantly influence the visual result~\cite{jankun2000spreadsheet}. Aspects such as the robustness against small perturbations of these parameters typically need to be continuously verified. While unit tests and other forms of testing are meant to fulfill a similar purpose, they are often only employed once a set of required features is clearly defined and already implemented in form of a prototype. In research, in particular, we often lack the detailed specification needed to clearly specify such tests from the very beginning. Based on previous observations~\cite{laramee2009using} and our own analysis of a typical visualization developer's workflow, we can identify a set of tasks that are regularly performed by the user and could benefit from additional support:\\

\noindent
\textbf{T1.} Compiling the visualization source code to investigate the visual result regarding functionality\\
\noindent
\textbf{T2.} Comparing visual results from different revisions\\
\noindent
\textbf{T3.} Understanding the impact of several implemented features on the visual outcome\\
\noindent
\textbf{T4.} Locating the parts of source code that are responsible for a visual feature or bug\\
\noindent
\textbf{T5.} Switching between source code states\\
\noindent
\textbf{T6.} Finding suitable input parameter values\\

These tasks are partially supported by common programming environments, but could in many cases be performed in a more efficient manner. \emph{Compilation} of the source code typically needs to be manually initiated, thus breaking the developer's focus on the implementation of the algorithm. Several visual results can only be compared by individually compiling and investigating the corresponding states, which have to be manually stored and restored.

Although the \emph{source code} creates the \emph{visual features} in the algorithm's result, these two aspects are never directly shown in \emph{relation} to each other. The source code that is responsible for a given feature can thus only be located by reading and understanding the source code. This task is only accelerated if the user knows in which order the features were coded. The same issue occurs when trying to find a bug in the source code. By the time the user manually initiates the compilation process, a lot of code might have been written and is suspect to having introduced the problem.

In order to \emph{compare} features and source code, the user must be able to switch between different states. This task is frequently performed to compare functioning to malfunctioning code, but not sufficiently supported by commonly used development tools. As bugs or other issues are often only discovered when inspecting the visual results, current revision management tools, which provide guidance by highlighting source code differences and listing manually-entered commit messages, are a suboptimal solution.

Finally, the analysis of \emph{input parameters} and their effects on different stages of the development process is only poorly supported by current tools and typically requires a cumbersome trial-and-error process.

\subsection{System Design}

\begin{figure}[tb]
 \centering
 \includegraphics[width=\columnwidth]{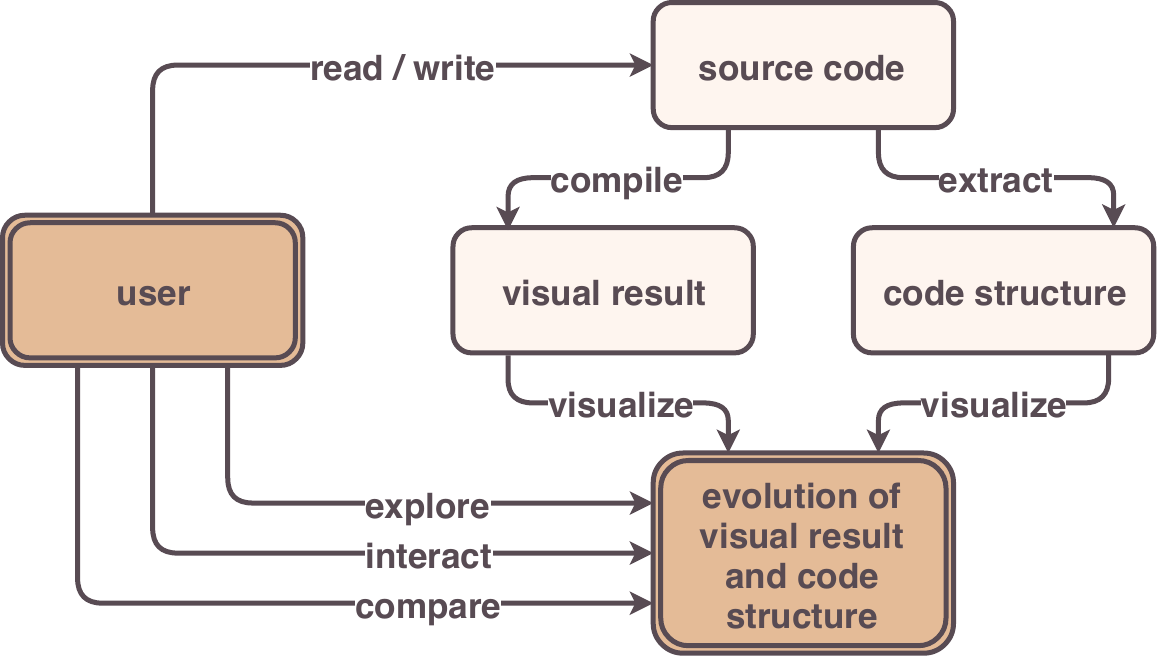}
 \caption{System overview. When a user writes visualization source code, its structure can be extracted and its execution provides a visual result. Following this procedure over time and combining both components into an interactive meta visualization, provides insight into the evolution of the developed algorithm.}
 \label{fig:system}
\end{figure}

Looking at the outlined shortcomings of commonly used systems with respect to the discussed user tasks, we are able to identify several key aspects that would greatly assist visualization developers:

\begin{itemize}
\item Visualization for investigating the visual results of several different states at the same time (T2, T3)
\item Visualization of the states' source codes in relation to their visual features (T3, T4)
\item On-the-spot switching between several states, visually guided by the states' visual results (T5)
\item Automatic compilation of source code in the background (T1)
\item Automatic and transparent storage of source code states (T2, T5)
\item Interactive elements for on-the-fly value definition of input parameters (T6)
\end{itemize}

The desire to provide an integrated solution that addresses these points lies at the heart of our system's design, which is illustrated in \autoref{fig:system}. We aim to assist the programmer in focusing on the implementation of the algorithm's features, understanding and comparing the impact of different choices on the visual result, and allowing for their comparison. We provide further support for the localization of issues and the automation of tedious tasks.

Our system was designed to flexibly support multiple back ends. It supports C++ and GLSL code as general-purpose programming languages, as well as Diderot~\cite{Kindlmann2016}, a domain-specific language (DSL) specifically designed for the development of scientific visualization algorithms. While C++ and GLSL have been utilized in many visualization applications for possible performance gains, DSLs provide benefits in usability and expressiveness, while reducing the complexity of the algorithm's source code with efficient syntax. Diderot specifically provides support for visualization-specific data and features, leading to compact and readable code. It further allows for parallel execution of the developed algorithm and thereby provides faster visual feedback. All examples and results in the remainder of this paper are either generated using Diderot, or a combination of C++ and GLSL code.

When analyzing the planned support functionalities to build a meta visualization which enables the exploration of a visualization's evolution, it becomes clear that our system requires access to the algorithm's source code and its visual result. Many visualization algorithms further depend on the definition of a proper parameter set to produce a meaningful output. We want to apply the same parameter changes to several different states of the source code evolution to compare their impact on the visual result. Therefore, input parameters need to be extracted from each algorithm to check which parameters can be applied. The toolchain of a programming language needs to meet two requirements to be supported by our visual analysis system:

\begin{itemize}
\item Facilities for extracting the visual outcome from the algorithm's execution
\item Definition of input parameters to steer the execution and influence the runtime environment of the algorithm via external tools
\end{itemize}

When these conditions are fulfilled, the programming language can be integrated into our system and benefit from all the additional functionality.

\section{Exploring Visualization Source Code}
\label{sec:methods}

\begin{figure}[tb]
 \centering
 \includegraphics[width=\columnwidth]{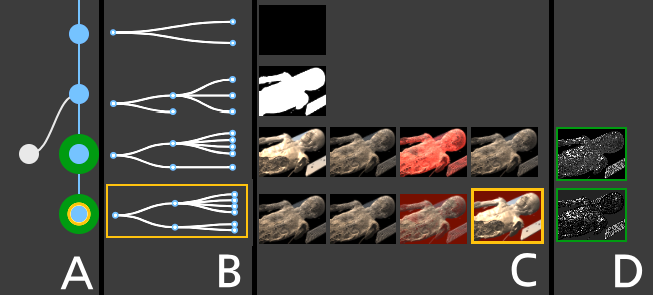}
 \caption{Meta visualization. (A) The revision tree of all stored states is shown as a node-link diagram, where every revision is represented as a node and a link between two nodes illustrates the evolution from the upper to the lower node. The number of links at the bottom of each node represent the number of branches evolved from this state. States and their links are colored in blue for the current branch and grey for other branches. The current state is highlighted in orange and collapsed nodes are highlighted in green. (B) The static scope tree outlines the source code's structure and is shown for each state of the current branch. Each node represents one structural block of the source code. A link between two nodes represents that the scope to the right is nested inside the scope to the left. (C) The comparison of visual changes is presented in a juxtapositional view. For each state of the current branch, the visual result created by the source code is displayed as an image from a user-defined viewpoint. For states which are collapsed in the revision tree representation, result images are positioned next to each other. (D) A comparison image shows the visual variance between all result images inside such a collapsed node.}
 \label{fig:visavis}
\end{figure}

In order to transcribe our main concept into a usable interface, we provide a meta visualization displaying the information of all visualizations created during the development process. All methods, their main ideas and comparison to alternative approaches, will be covered in the following. The resulting meta visualization, produced from an exemplary development process, is depicted in \autoref{fig:visavis}.

\subsection{Automatic Revision Management}
\label{revisions}

A central goal of our approach is to enable and support the comparison of several states during the development of a visualization algorithm. Many of our design choices draw inspiration from the VisTrails~\cite{Bavoil2005} approach for visualization pipelines. In order to access a given revision and all its information at a later point in time, it needs to be stored. Many applications make the user responsible for storing their progress by enabling them to manually save the current state. This approach only provides access to the last development state and is prone to system failure, whereas smarter systems create automatic backups. In practical software development, version control systems like Git~\cite{Git} or SVN~\cite{SVN} are typically employed as they store the entire revision history. However, the process of storing an individual revision still needs to be triggered manually.

We internally utilize a version control system (Git) and draw inspiration from modern tools such as Google Docs that automatically track the version history of a document without user intervention. Instead of constantly interrupting their workflow, users can simply focus on the development process. We automatically track and store every code edit and create a new revision whenever the corresponding source code compiles successfully. All automatically stored states are displayed to the user in a node-link diagram to keep transparency over the process and visualize the development progress (\autoref{fig:visavis}A). Each node represents a development state and a link between two nodes describes the evolution from one state to the other, or, in other words, that the successor node (child) was created by modifying the preceding node (parent). Interacting with the visualization allows for intuitive switching between development states and further eases the comparison task. If the user switches to a previous state, which by definition already has a child, and continues coding, a new branch is created and the compiled state is represented by a new node.

\subsection{Visualization of Algorithm Evolution}
\label{sec:ast}
\label{sec:metavis}

After having described our visualization of the revision tree, we will in the following describe the other visual components which compound our meta visualization: the visualization of source code structure, the visual result, and the difference image.\\

\noindent\textbf{Result Image}

\noindent
Writing source code is an error-prone task and mistakes in the program can result in failure or incorrect visual results. The visual output can serve as a compact descriptor of the functionality and features integrated into the given source code state. It can be utilized to easily identify different feature sets and to add, remove and exchange them. Furthermore, having an overview over existing features can reduce redundant code and effort in reimplementing features. For each compilable revision, we execute the program and display the result image (\autoref{fig:visavis}C) aligned with the corresponding node of the revision tree and a representation of the generating source code (\autoref{fig:visavis}B).\\

\noindent\textbf{Source Code Structure}

\begin{figure}
\centering
\begin{minipage}[b]{.47\columnwidth}
    \begin{subfigure}[t]{\columnwidth}
      \centering
      \includegraphics[width=0.75\columnwidth,height=2.4cm]{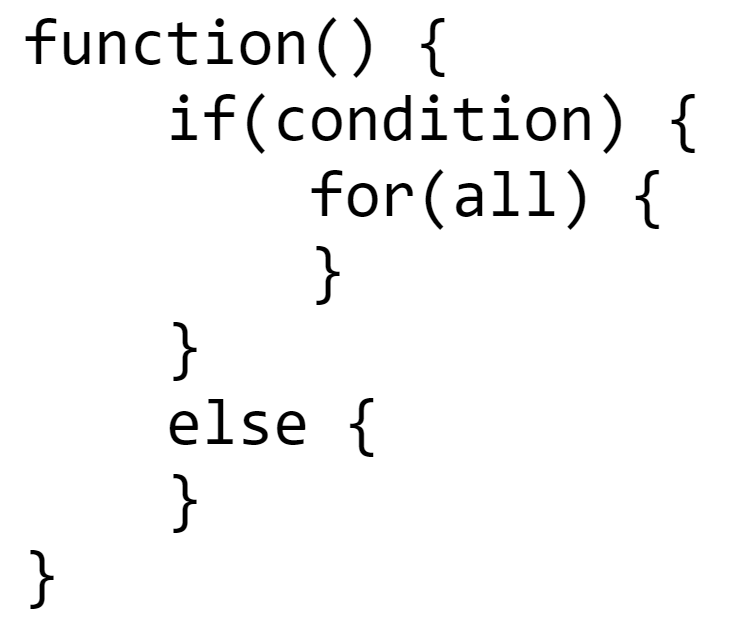}
      \caption{}
      \label{fig:sub1}
    \end{subfigure}\\
    \begin{subfigure}[b]{\columnwidth}
      \centering
      \includegraphics[width=0.7\columnwidth,height=1.6cm]{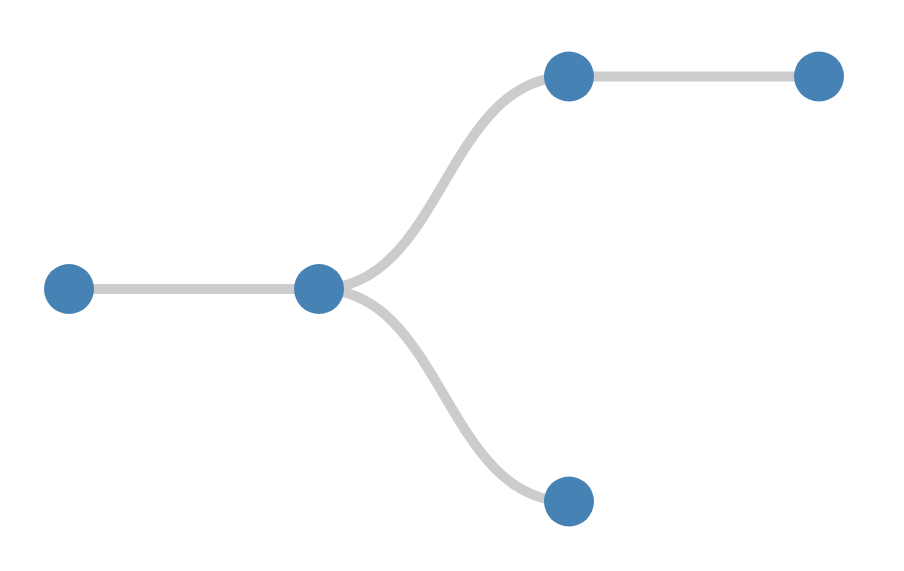}
      \caption{}
      \label{fig:sub2}
    \end{subfigure}%
\end{minipage}
\hfill
\begin{minipage}[b]{.47\columnwidth}
    \begin{subfigure}[b]{\columnwidth}
      \centering
      \includegraphics[width=\columnwidth,height=4cm]{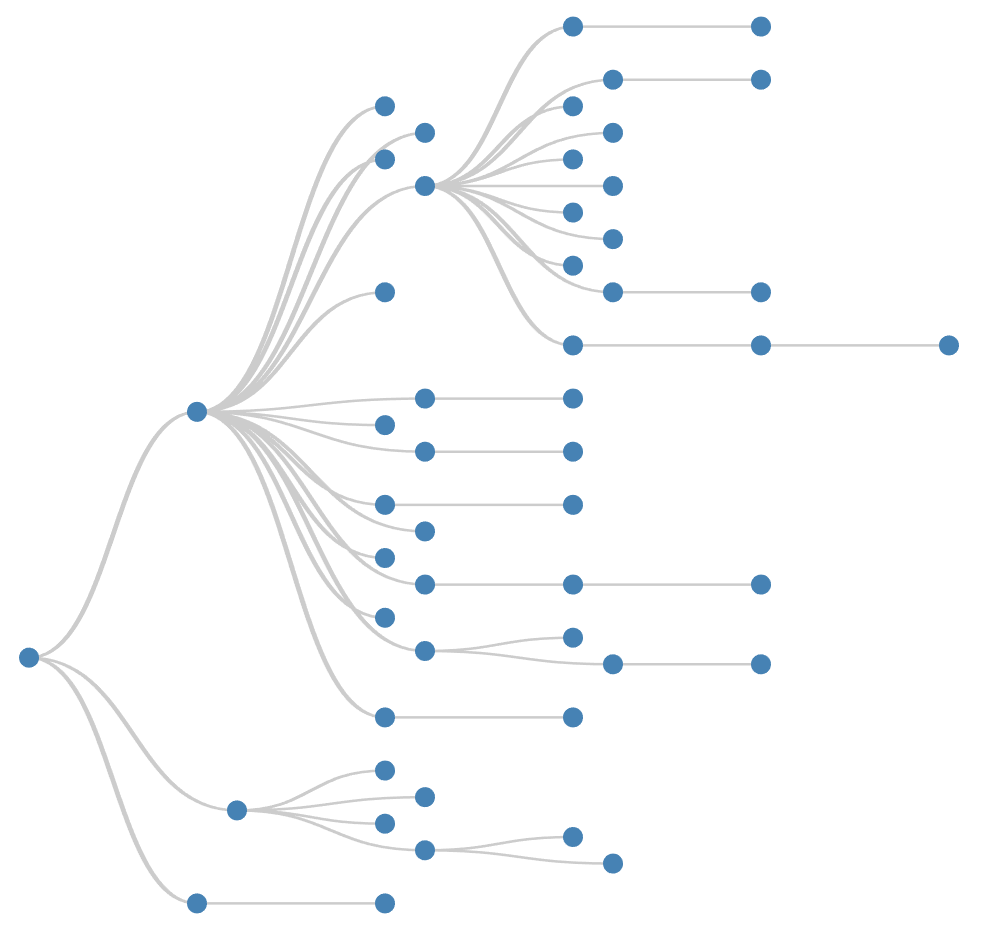}
      \caption{}
      \label{fig:sub3}
    \end{subfigure}
\end{minipage}
\caption{Static scope tree. (A) Pseudocode. (B) SST extracted from A. (C) SSTs extracted from three source code files (from top to bottom: C++ source, GLSL, C++ header) and combined under a mutual root node.}
\label{fig:SST}
\end{figure}

\noindent
While there are many different techniques for visualizing source code structures, as discussed in \autoref{sec:related}, we focus on its outline in the form of a static scope tree (SST). This representation highlights the nesting scopes (block structures) of the source code and is sufficiently compact while still conveying the main structural aspects of the code. In both programming languages featured in this paper, nesting scopes are defined by opening and closing curly brackets. The SST is visualized by a node-link diagram (\autoref{fig:SST}B), where every node represents a code block and a link exists between two nodes, if one block is nested within the other. The tree depth represents the nesting level. If several source code files exist, the SST is computed for each of them, and their root nodes are linked as children to a new root node which represents the files' directory (\autoref{fig:SST}C). While the SST provides a high-level overview for source code comparison, the actual differences between two given revisions are computed and visualized in a tooltip to highlight their code changes in detail.\\

\renewcommand{\thesubfigure}{\Alph{subfigure}}
\captionsetup[subfigure]{labelformat=simple}

\noindent\textbf{Compression}

\noindent
Since every compilable development state is taken into account, the meta visualization can become very large. We therefore compact the revision tree based on a predefined comparison measure to enable the representation of several successive states within a single collapsed node. A node and its parent are represented by a single node, if both states are equal with respect to the chosen comparison method. The representing node inherits the children of all collapsed nodes and the visual results of all states being combined in such a node are displayed next to each other. If desired, the user can manually expand and collapse the node for detailed investigation. In a first iteration, we define the comparison measure as the similarity in source code structure. We assume that a structural change describes a major change in the source code. All successive nodes that represent states with the same source code structure are bundled in our default abstraction. The corresponding SST is only displayed once per bundle. One can easily think of other possible comparison measures, like defining thresholds for visual differences in the result image, number of code changes or performance differences. The question of which compression method is best suited for the exploration of visualization source code evolution is left open for future research. Alternatively, letting the user tag states of interest for comparison could be a good alternative to automatic compression approaches.\\

\noindent\textbf{Result Comparison}

\noindent
The identification of subtle visual differences between the results of multiple development states can be challenging. For this reason, we provide an additional comparison view (\autoref{fig:visavis}D) that shows the per-pixel variance of the individual results for states collapsed within a single node. The resulting image is black if all compared images are exactly equal. The higher the color difference between the images in a certain location is, the brighter is this part of the computed image. Our approach of showing visual results next to each other and visualizing the small differences in an additional view benefits from the advantages of both a juxtaposition and explicit encoding~\cite{Gleicher2011}. It enables the user to identify visual differences on both the large and small scale. Combining the visual differences with the changes in source code and aligning them along the algorithm's evolution, provides the visualization developer with an enriched insight into the process that would require significant effort to be achieved with common development tools.


\subsection{Parameter Management}
\label{sec:viewer}

Many visualization algorithms define several input parameters, which can significantly alter the visual result. Investigating the impact and interplay of individual parameters on different stages of an algorithm is a non-trivial task. Simple systems supporting this task can modify certain parameters on the application level and visualize the result directly. Some of them store the visual result of several parameter settings to ease the task of comparing different parameter sets. When the user wants to test if the change of a parameter has different impact on individual features of their implementation, these systems provide little support. The user would need to run the available tool for each state of interest, change the parameter settings in every instance, and compare across these instances.

\begin{figure}[tb]
 \centering
 \includegraphics[width=\columnwidth]{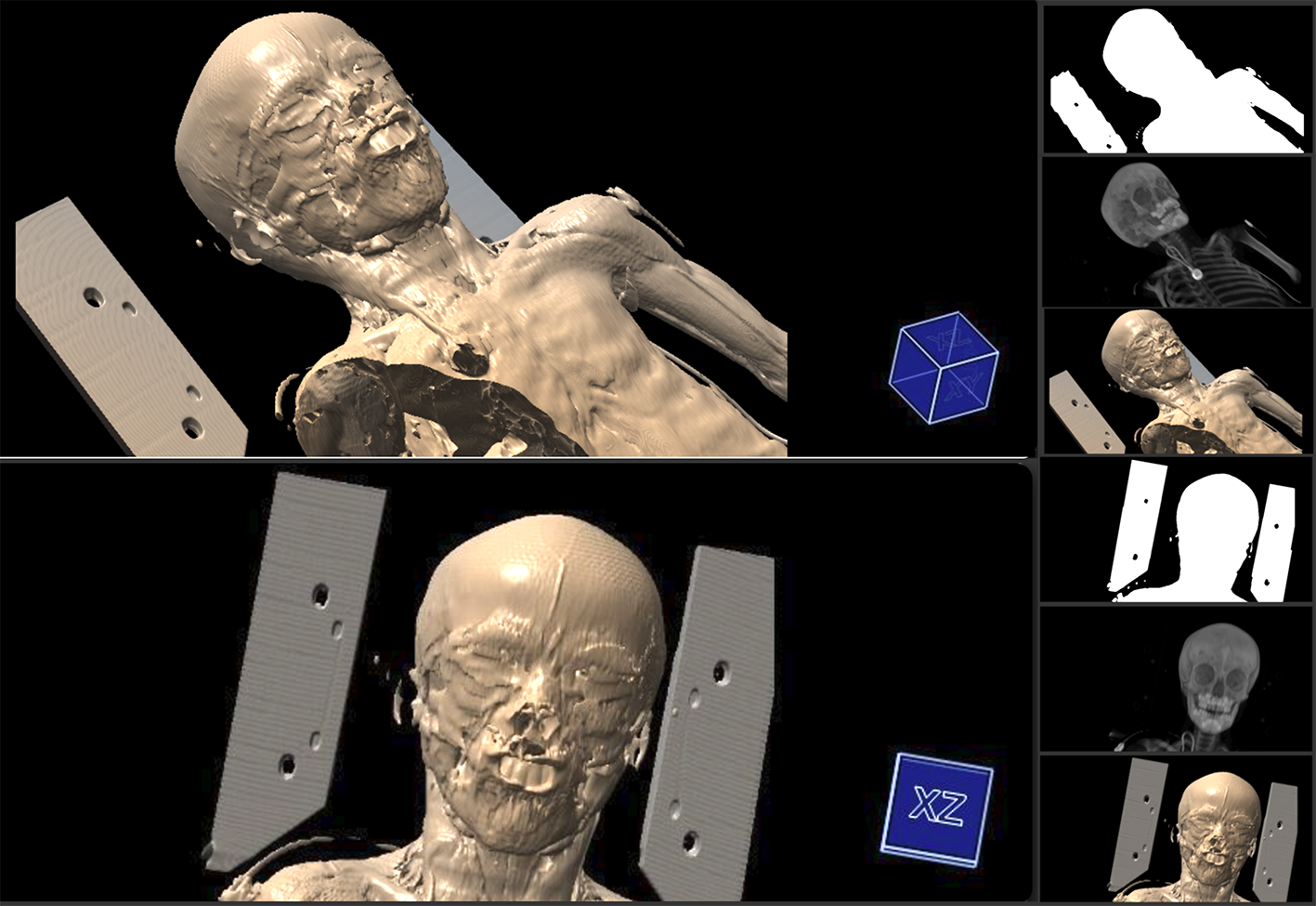}
 \caption{The Live View enables direct user interaction including rotation, transformation, and scaling to investigate the visual result of a given state in detail. These interaction functionalities are automatically integrated, as soon as the developer utilizes the predefined keywords. The newly chosen viewpoint is automatically applied to all other visual results to enable consistent comparison between states.}
 \label{fig:viewer3d}
\end{figure}

Our system provides the possibility of freely exposing all input parameters of a state and mapping them to individual type-specific elements of the interface which provide convenient interaction facilities for parameter changes similar to common property sheets, using simple syntactic constructs specific to the host language. Based on this mechanism, we are further able to provide more intuitive, specifically tailored interaction elements for common parameters in the visualization domain. The virtual camera is one of the most common parameter sets in three-dimensional visualization, typically specified using an interaction technique such as the Arcball~\cite{Shoemake1992}. It controls the rotation of the camera around an object and can be easily expanded to support translation and scaling. We integrate this well-known technique via a plugin mechanism into our system. The plugin defines three keywords for the camera's position, look-at point and orientation. When the user utilizes these keywords as names for the corresponding parameters in the algorithm's source code, they are automatically coupled. We display an enlarged version of the current state's visual result for detailed exploration, that can be seen in \autoref{fig:viewer3d}. Interacting with the view automatically runs the algorithm with the new parameter set and updates the result view on-the-fly. This approach expands the visualization algorithm by a direct, interactive component without any extra effort. Our system provides the ability to automatically and transparently make other such interaction facilities available for developers to directly integrate into their algorithm.

To support the task of comparing the visual outcome of different parameter settings on several states, we perform the parameter change on all visual representations of states within the current development branch. This means that, for instance, moving the camera of the current state will automatically update all other states to the same camera position. The user can then explore whether states with different feature sets undergo different degrees of visual change. If a parameter is not present in a certain state, it is simply ignored. This semi-automatic approach enables the user to investigate the impact of a parameter change across all implemented feature sets. It further provides insight into the sensitivity of different features to certain parameter changes and into the robustness of investigated parameters in terms of their compatibility with multiple features.

\subsection{System Interactions}

\begin{figure*}[tb]
 \centering
 \includegraphics[width=\textwidth]{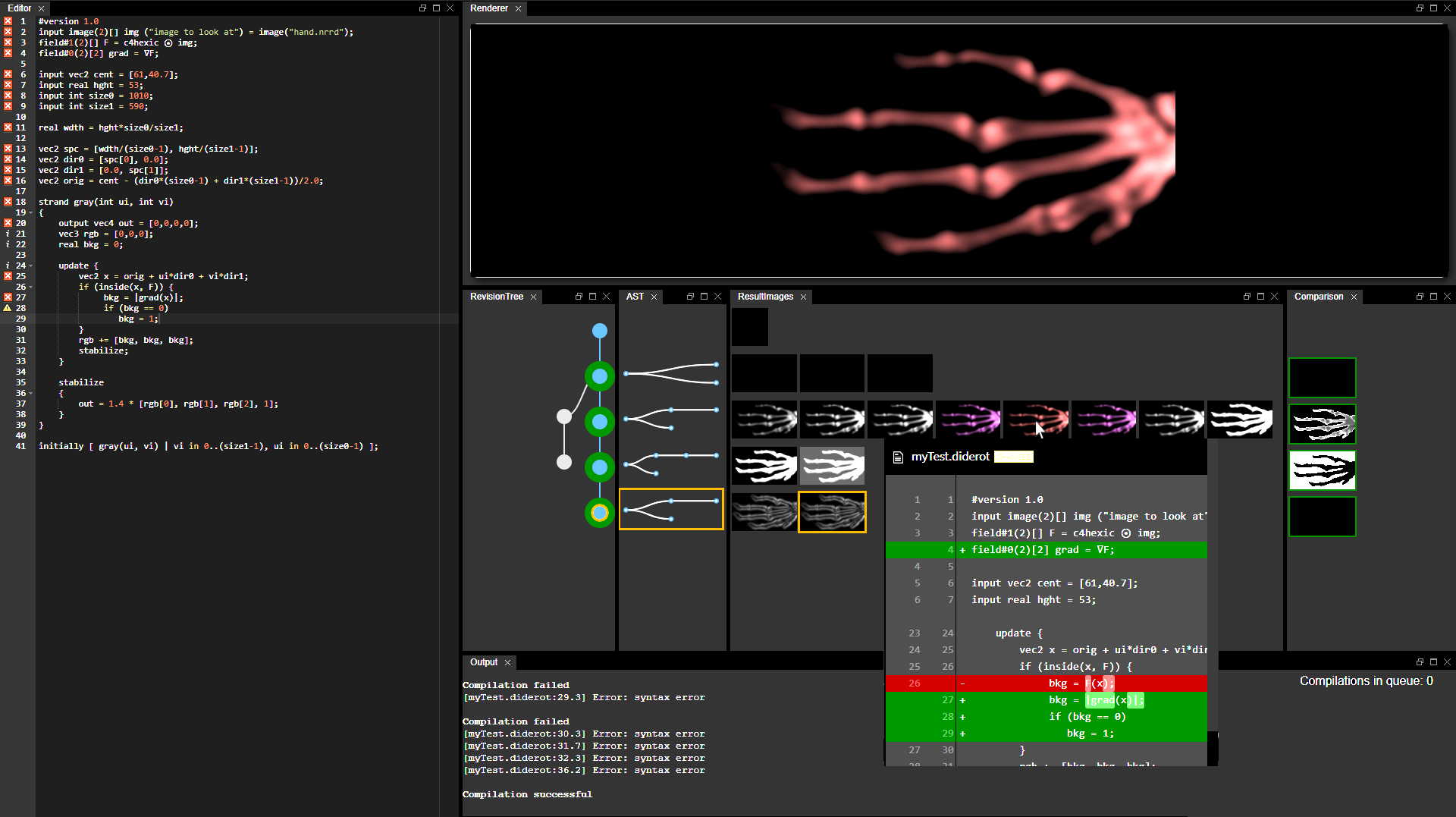}
 \caption{Graphical User Interface. The interface of our application is inspired by common Integrated Development Environments. Every view can be individually positioned and resized based on the user's preferences by easy drag\&drop interactions. The code editor is shown to the left and includes the source code of the current state's visualization algorithm. The whole evolution of the algorithm, up to the current state (highlighted in orange), is presented in our meta visualization. The top right view would normally display the visual result of the current state, but since the user hovers over one of the states in the meta visualization, its result is shown instead, enabling easy comparison between the two. Additionally, the source code differences between the hovered and the current state are directly visible in the overlaying tooltip. The bottom right view displays the compilers output after each compilation attempt.}
 \label{fig:interface}
\end{figure*}

Although all described methods can be utilized by themselves, they further benefit from being visualized next to each other and interconnected by user interactions. To provide a better picture of the possible interactions, we present an overview of our graphical user interface in \autoref{fig:interface}.

The visualization of our revision management system, the evolution of source code structure, and the evolution of visual results all communicate progress over time. Their respective components are each related to certain states of the development process. For every compiling state of source code, a node in the revision graph is displayed, the source code's structure is visualized, and a view of the visual result is shown in an image. All three visual representations are aligned along a common time axis. This approach further emphasizes the evolution from one revision to the other in a clearer manner by following the time axis from top to bottom, as shown in \autoref{fig:visavis}.

We display additional overlays for a more convenient comparison of several states. When hovering over a result image, the system displays a tooltip which includes all source code differences between the hovered and the current state. To easily find these differences in the code editor, all lines of code that need to be removed and added to get from one state to the other are highlighted and their line numbers are displayed. This interaction mechanism allows for an in-depth inspection of the source code changes. The current state can be easily switched by clicking either on the result image of a state, or its node representation in the revision tree. Both elements, as well as the corresponding static scope tree, are highlighted, so that the user always knows which state they are currently working with.

Since the revision tree collapses nodes if their source code has the same SST, it is necessary to provide the user with tools to investigate all available states. A right click on one of the collapsed nodes will expand all hidden nodes and branches. In order to keep the visual components in alignment, the SSTs and result images of the currently shown branch will be repositioned. Since such a displacement might confuse the user's mental map, hovering over a SST highlights all result images that share the respective SST. Following the same principle, hovering over a comparison image highlights all result images that were taken into account to compute it. The expanded nodes in the revision tree can be manually collapsed again by performing a right click on any of the nodes which share the same SST.
Clicking on a node in the revision tree which is not within the current branch will switch branches and update all views to visualize the meta information of the selected branch. All nodes of the currently visualized branch are marked in blue in the revision tree, while all other nodes are colored in grey. It is important to keep these non-active branches within the visualization, because otherwise there would be no interaction available to activate them again.

When hovering over a result image, it is enlarged for easier inspection. It provides the user with a preview of what is to be expected, when switching to that state. Continuous alternation of hovering over two images allows for easy comparison and detection of visual differences. To detect even the smallest differences, the comparison image is also enlarged when being hovered.

\section{Implementation}
\label{sec:implementation}

Our application is divided into a server and a client component, each benefiting from different programming languages and execution environments, whose technical details will be briefly outlined in this section. We provide a short description of the components' functionality, communication, and the events being triggered by user interaction.

When setting up a machine for software development, many tools, like compilers, libraries, or an IDE, need to be installed. The chosen tools and developed source code often vary based on the underlying operating system or hardware used. Finding the right tools for a given task can be very time-consuming, especially for beginners in software development. We therefore chose a client-server architecture to relieve the user from setting up the environment needed for compiling and executing the source code on the client side. It enables us to run a large amount of compilations in the background without affecting the user's workflow. Our solution can therefore even be used on low-end clients.

In order to make our application widely available, we implemented the client side as a web application. In addition to standard web technologies (HTML5, CSS3, and JavaScript), we utilize the Ace Web Editor~\cite{lib:ace} as environment for writing visualization source code, and D3~\cite{Bostock2011} to create node-link diagrams within the meta visualization. The window layout is based on Golden Layout~\cite{lib:golden} and is inspired by common Integrated Development Environments enabling the developer to add, delete and move modules based on their personal preferences. The visual presentation of source code differences is handled by the diff2html library~\cite{lib:diff2html}. Our server is written in C++ and handles the compilation of source code, storage of states, visualization of results, and the communication with web clients. The communication channel between client and server is based on the JSON-RPC protocol. Data is requested via AJAX calls and exchanged in the JSON format. We further use Git~\cite{Git} as the version control system by utilizing the functionality provided by the libgit2 library~\cite{lib:git2}. The current version of our application supports Diderot~\cite{Kindlmann2016} as a programming language specifically designed for the visualization domain, as well as C++ and GLSL. While Diderot code is commonly manageable within a single source file, C++ and GLSL code can become quite large. We provide multi-file support to split up source code into several files. In order to support additional programming languages, the system requires access to a corresponding toolchain, the definition of commands to build both a library and an executable from the source code, and access to the result images. Everything else is handled automatically, e.g., error messages from the toolchain are forwarded to the client and source code is stored in Git.

In the beginning of the development session, the user chooses a programming language. Based on this choice, the server prepares the appropriate compiler and runtime environment to compile and execute all incoming source code. When the user writes code in the editor, our system waits until the user stops typing for a certain amount of time (default 1.5 seconds), before sending the code to the server and compiling it to an application. The static scope tree is automatically extracted during the build process. If the compilation was successful, the source code is stored in the Git repository and the compiled executable is cached for future execution. The visual result is sent to the client along with the current state of the revision graph and the SST of the source code. If the compilation fails, an error message is sent to the client. When the user interacts with the live view, the visualized state is rerun with the new parameters on the server side and the results are sent to the client. The same parameter set is used to run every stored revision of the current branch and thereby updates all visual results, which are then sent to the client in an asynchronous manner. In order not to introduce any decrease in performance compared to common development environments, where the user would manually start the compilation process, we prioritize the compilation and update of new source code over older revisions. We realize this by utilizing a priority queue, where compilations are assigned the highest and updates the lowest priority. When two operations in the queue have the same priority, we perform the latest request first. For further performance improvements, older compilation requests in the queue can be dropped when a newer revision was successfully compiled. Parameter updates across the revision tree are only performed when the hardware resources are available. If the user wants to switch to another state in the revision graph, the server finds the given state by its unique ID in the Git repository and sends the source code to the client. In the same way, source code differences are computed between two revisions on the server side and sent to the client for display. When managing several source files, the user can switch between these files by clicking the corresponding tab in the code editor. We check if the file contains the content of the currently selected state and perform an update if necessary.

\section{Usage Examples}
\label{sec:results}

Having described the individual components of our system and how their capabilities and interconnections employ our conceptual methods, we now want to illustrate the system's advantages over commonly used systems in real-world scenarios. As interactive processes are inherently difficult to capture in still images, we encourage the reader to also refer to our supplementary demonstration video. We present the implementation of two usage scenarios of our approach: a three-dimensional flow visualization and a visualization with stylized line primitives. We highlight the specific integrated features and discuss how they enhance the user's experience during the development process.

\subsection{Flow Visualization}

\begin{figure}[!tb]
 \centering
 \includegraphics[width=\columnwidth]{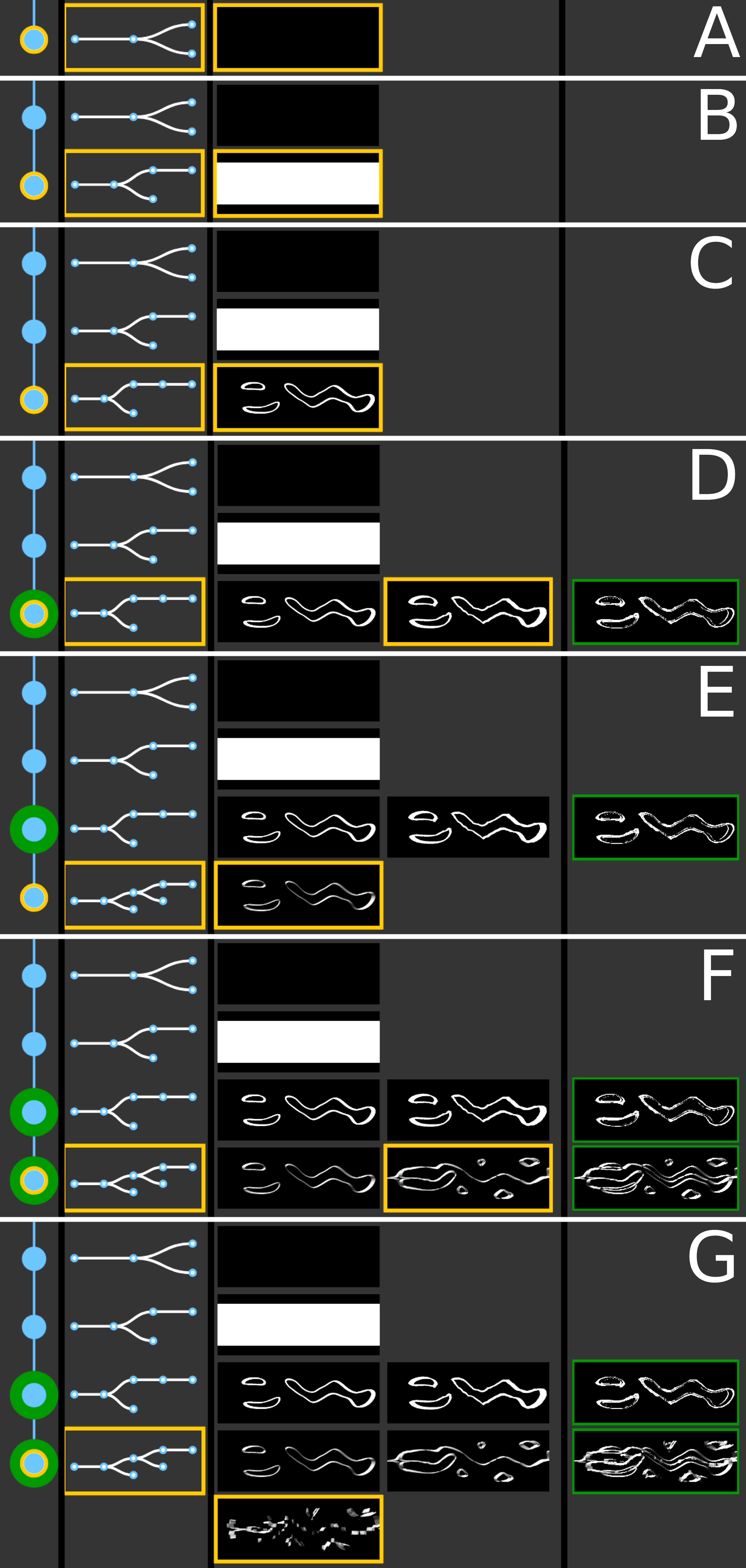}
 \caption{Visualization of a visualization source code evolution. (A) Initial state. (B) Implement ray tracing and find volume. (C) Implement flow visualization based on flow magnitude. (D) Test trilinear interpolation instead of B-spline interpolation. The comparison image highlights minor differences. (E) Implement shading. (F) Change the flow feature from flow magnitude to extremum lines. (G) Test trilinear interpolation instead of B-spline interpolation.}
 \label{fig:result-story}
\end{figure}

The user's task is to visualize specific properties of a vector field, starting from flow magnitude, over extremum lines, to normalized helicity. Since the given vector field has three spatial dimensions, the visual result shall be three-dimensional as well, so a ray casting algorithm will be used. In order to easily follow the development process and to get a better impression of how our system's support functionality looks like, we display all intermediate results of the now following description in \autoref{fig:result-story}.

The process starts with an initial state which only consists of Diderot's boilerplate code and a black image as its output (\autoref{fig:result-story}A). With the three-dimensional result in mind, we add camera parameters to the algorithm, cast a ray at every pixel of our camera's resolution and display data in white, when it is hit by a ray (\autoref{fig:result-story}B). At this point already, by utilizing the predefined parameter names for the camera's position, look-at point and up-vector, our application will automatically assign values to these parameters, allowing the user to zoom, pan, and rotate around the data set and select the best possible viewpoint.
In order to gain a better impression of the flow, we compute the flow's magnitude and only display it for values above a constant threshold (\autoref{fig:result-story}C). At this point, we start experimenting with the interpolation kernel, which is responsible for creating a continuous field from our given input data (\autoref{fig:result-story}D). Diderot has convenient built-in mechanisms to change this reconstruction method. Investigating the result images clearly shows a much smoother flow when using B-spline interpolation compared to trilinear interpolation. The comparison image highlights subtle differences to the user. Now that the flow is already visible, we can improve shape perception by adding a light source into our environment and performing gradient-based shading (\autoref{fig:result-story}E). Since the basic visualization algorithm is functioning, we can start to integrate additional flow features. Visualizing surfaces around extremum lines and vortex structures is easily done by modifying the computations over the corresponding vector field (\autoref{fig:result-story}F). After these additions, we notice artifacts when experimenting with different interpolation methods (\autoref{fig:result-story}G), both in the case of trilinear interpolation and even when using Catmull-Rom splines. By hovering over the images of previous states, a tooltip reveals our last code changes and thereby highlights the different vector field computations. 
We check the maths and consider our algorithm as correct. The comparison to even earlier states ensures us that no drastic changes were made and the issue must be related to the interplay between flow features and the reconstruction kernel. Also, our basic algorithm seems to be correct, since the visualization of the flow magnitude does not show any problems. The only difference and possible cause of the issue in this scenario therefore seems to be the gradient computation. If we, as a user, had the necessary background knowledge, we would know that trilinear interpolation does not provide the continuity necessary to compute appropriate derivatives. Our application was therefore able to reveal the complex correspondence between interpolation kernels and gradient computation by displaying the visual results and source code differences in an explorable manner. Exposing such high-level relationships without explicitly knowing the reasons for their existence, requires the human-in-the-loop for further sense-making, but provides a hint to in-depth understanding of the algorithm's underlying functionality. The user can now further compare the results of functioning states and visual differences in all the implemented functionality, choose the most suitable version for their task and add further features if necessary. When investigating the evolution of the static scope tree over the whole process, it is noticeable that the source code structure expanded until shading was implemented, but stayed constant afterwards. This provides us with an indication that all main features being considered during this session require the same fundamental code structure in order to function, but mainly differ in the mathematical formulae being computed over the flow field.

\subsection{Stylized Line Primitives}

\begin{figure}[!tb]
 \centering
 \includegraphics[width=\columnwidth]{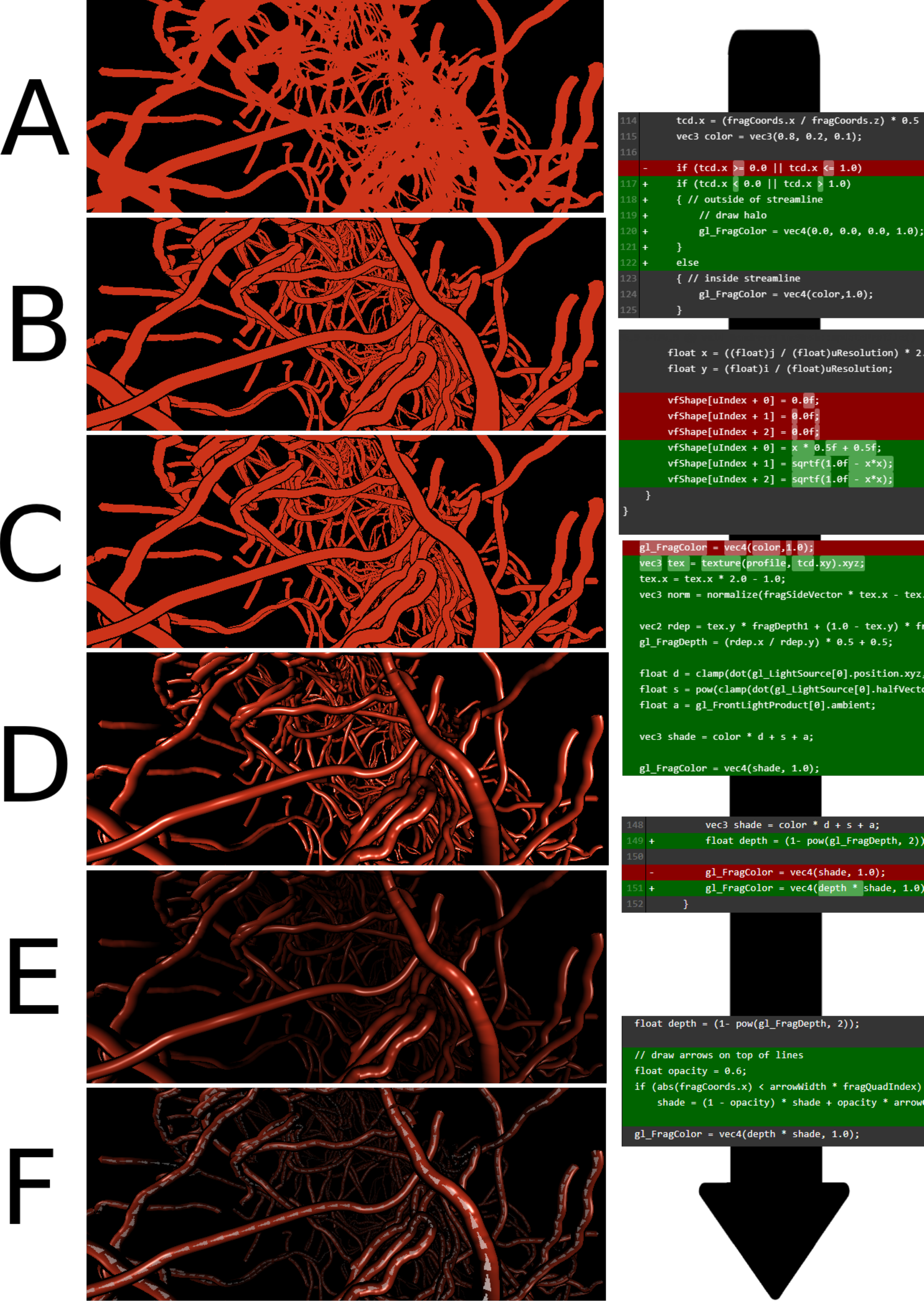}
 \caption{Visualization with stylized line primitives. Only extracts of source code differences are shown. (A) Implement basic rendering of line primitives. (B) Add a black halo to all lines. (C) Change the three dimensional data profile of lines. (D) Implement shading. (E) Implement depth enhancement. (F) Draw arrows on lines, to describe the flow direction.}
 \label{fig:result-story-cpp}
\end{figure}

In our second example, we focus on efficiently drawing lines in a three-dimensional scene, by rendering them in 2D, but shading them as if they were 3D tubes~\cite{stoll2005visualization}. Additionally, several styles can be applied to the lines to represent different features in the data. It differs from our previous example by being implemented in C++ and GLSL, and thereby utilizing multiple files -- a C++ header, a C++ source and a GLSL file. Furthermore, the underlying source code and the corresponding source code changes contain many more lines of code (665 LOCs for the final state) than the previous example being written in a domain-specific language (72 LOCs for the final state). This scaling in data size comes with increased compilation times, bigger static scope trees and larger tooltips. The evolution of the algorithm's development is shown in \autoref{fig:result-story-cpp}. Since some of the source code changes being made are quite extensive, we focused on the most salient parts.

We skip the initial development involving data handling and the setup of the rendering process to focus on the evolution of the visualization techniques. We therefore start out with source code which already renders a red quad strip for each line in the data, following its curvature (\autoref{fig:result-story-cpp}A). All quads are, based on the user's viewpoint, computed and rendered in every frame. The definition of specific parameters in the code automatically provides the same interaction handling for changing the viewpoint as demonstrated in the previous example. In order to visually separate the lines from each other, we plan to add a halo to each line. We increase every quad's width by a margin and check for each fragment being drawn, if it lies within the line's width or exceeds it. If it exceeds the line's width, we paint the fragment black as a halo, otherwise we set its color to red (\autoref{fig:result-story-cpp}B). The visual difference becomes immediately apparent and provides us with a better understanding of the lines' paths and depths. In the next step, we want to give our lines a three-dimensional shape, which is why we define a texture on top of the quad, that contains the side vector, up vector and depth correction factor at each point (\autoref{fig:result-story-cpp}C). While all other source code changes are made in the GLSL file, this change is taking place in the C++ source file. Although the profile texture is now correctly set, the visual result appears to be the same and a look at the difference image confirms that they are identical. The change is not visible, because the texture is not yet applied and no shading is implemented. We implement Phong shading and now the lines appear as shaded three-dimensional tubes (\autoref{fig:result-story-cpp}D).  Unfortunately, it is not as easy to perceive the depth of the tubes in the scene, especially if they do not intersect in the result image. We therefore improve the rendering by integrating depth enhancement (\autoref{fig:result-story-cpp}E). Interacting with the visual result and comparing it to the previous states demonstrates the advantages of the newly implemented feature. The tooltip shows that this feature only consists of a few lines of code and can therefore easily be added to other visualization algorithms as well. As a last step, we draw arrows on the tubes to provide the viewer with the information on the lines' flow direction (\autoref{fig:result-story-cpp}F). The static scope tree of the final state is shown in \autoref{fig:SST}C. It provides an overview over the source code by clearly showing that it consists of three source files (because the root node has three children), where the C++ header file only consists of a single structural block, the GLSL file is a bit more structured, and the C++ source file contains the most structuring elements.

Looking at the performance of our system in this example, the compilation of source code took around 8 seconds, if it was successful. If it fails, the compilation process is canceled much faster and provides a description of the error. Since the compilation takes place on the server side, the user does not experience any slowdown in their coding workflow. Changing the viewpoint on the visual result took around 100ms per image. This means that when the user is only interested in the most current history (last ten states), their result images can be computed within a second, since images are updated starting from the latest state. Updating images for 100 states would take around 10 seconds, with the client view being progressively updated as new images become available. The number of triggered compilations and created states vastly vary based on the given task and the user's typing behavior. While typing, 10--15 compilations can be triggered every minute, of which 4--5 will be executed, while others are ignored in favor of newer revisions. Most of the time, only one of these compilations is successful and creates a state in the visualization. Users ended up with 30-50 states when working on this example.

\bigbreak

We have shown that our novel approach is able to reveal complex correlations between the visual result and the underlying source code of a visualization algorithm. It provides the user with direct feedback, enabling them to discover implementation problems as soon as they appear. It opens up new ways of comparing visualization algorithms by utilizing novel viewpoints onto the available meta data and thereby generates greater knowledge about the algorithms themselves. Revealing these relations between algorithmic techniques, mathematical formulas, implementation in source code, and visual outcome can greatly benefit the task of comparing visualizations on all these levels and be especially beneficial for teaching visualization algorithms to students.

\section{Evaluation}
\label{sec:evaluation}

Given the subject matter of our work, we performed our evaluation in two rounds and in the form of an expert review~\cite{Tory:2005:EVD}. In the first round we gathered qualitative feedback on the functionality and usability of our presented framework using the Diderot language. We selected 4 experts from academia with different specializations in the field of visualization. All participants rated their expertise in visualization between knowledgeable and professional. They had extensive experience in writing visualization source code (on a monthly to daily basis), but none of them were intrinsically familiar with Diderot. Based on the initial feedback, we improved our system to support C++ and GLSL code, handle multiple source files, and cache compiled source code for faster state switching. We then conducted a second round of evaluation to gather feedback on the improved state of the system by selecting 4 new experts with similar experience and an example using C++ and GLSL. All participants were familiar with different existing toolchains, ranging from C++ IDEs to web based development platforms and we asked them to assess our system in the light of their experience. None of the experts are co-authors of this paper, participated in the development of our system, or had used it prior to the review.

Following the guidelines of Tory and M\"{o}ller~\cite{Tory:2005:EVD}, the evaluation was split into several sessions, interviewing one participant at a time and following the same protocol: At first each participant filled out a sheet of information describing their personal background and expertise in visualization algorithm development. They were further introduced to the concepts of our approach and the application's functionality. All interaction possibilities were summarized on a two-page handout given to the experts. We not only wanted to assess the functionality of our application, but also the meta visualization's ability to communicate information about the visualization algorithm's development process. We therefore provided the participants with the visualization examples from \autoref{sec:results}, depicted in \autoref{fig:result-story} (round 1) and \autoref{fig:result-story-cpp} (round 2). Based on this initial state, the participants were asked to perform a set of simple tasks and verbalize their thoughts and reasoning behind their actions (think-aloud protocol). They were allowed to ask questions about the tasks and application at any given time. The tasks were performed without time limit and designed to encourage the exploration of all aspects of our system. Tasks like \emph{"Find a state with a bug"} and \emph{"Figure out, what the reason for the bug is"} required the user to analyze the whole development process on the visual and algorithmic level. Asking the participants to \emph{"Change the code, so that it produces a different output"} not only made them actively develop within our framework, but created personal and for us unexpected results that led to diverse usage of the provided system. When the users completed the tasks, they received a 28-statement questionnaire to answer on a 5-point Likert scale. It covered the difficulty and suitability of the given tasks, general functionality of the application and assessment of individual components (e.g. \emph{"I found the abstract code view was giving a good impression of the source code's complexity."}). It further included the ten statements of the System Usability Scale (SUS), which allows for the uniform comparison of different systems in terms of usability and user satisfaction. Last but not least the participants were asked to openly summarize what they liked/disliked about the system, what they would like to change about it, and how they rate its practical impact. The participants' feedback to each question, except for the SUS (Q1-Q10), is illustrated in \autoref{fig:evaluation}. All the answers can additionally be found in the supplementary material.

\begin{figure*}[tb]
 \centering
 \includegraphics[width=\textwidth]{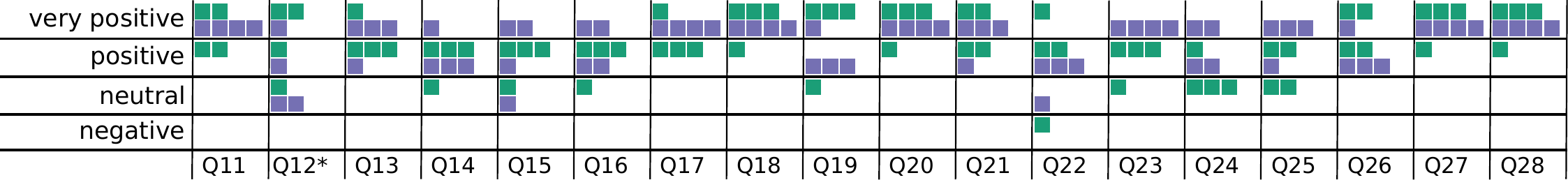}
 \caption{Illustration of the participants' answers to statements of our questionnaire on a 5-point Likert scale. Very positive feedback was reflected by either strong agreement to a positively formulated question, or strong disagreement to a negatively formulated question (marked by an asterix). Very negative feedback was not given by any participant. The color reflects the participants group, green for the first round and purple for the second round.}
 \label{fig:evaluation}
\end{figure*}

While we cannot discuss all the feedback in detail here, we want to highlight some of the most interesting insights gathered through our evaluation. All participants rated the system's functionality as useful (Q11) and rated the availability of such a system as beneficial for their work or studies (Q13). When assessing the availability of other tools with similar functionality (Q12), participants named Git and Shadertoy (in round 2), but also mentioned the superiority of our approach given the extended functionalities it provides. The participants mentioned several times that they would like to use our system as a rapid prototyping tool, or as a platform for teaching visualization algorithms to students, but that larger software projects would require additional features to support the development of other parts than visualization or rendering modules.

Looking at the different components of our system, the result view was easy to understand (Q18), easy to use (Q19) and helpful in completing the given tasks (Q20). The convenience of seeing all the visual outputs, investigating the source code differences by simply hovering over them and the intuitiveness of switching back and forth between these states, were highly praised by several participants. The fact that changing the viewpoint in one state also updates the viewpoint in all other result images, was recognized as being helpful for being able to compare the results to each other. One participant raised the reasonable doubt that finding a certain state within the result images after its parameters were changed might be difficult, since it is then harder to recognize as being the same state. The revision tree gathered a more mixed feedback, with most experts finding it easy to use (Q23) and, to a lesser degree, finding it easy to understand (Q22). Several participants mentioned that the knowledge of how version control systems work is beneficial in understanding the visualization and using it. We believe that most of the problems with this view originate from the fact that several branches can be collapsed in the same node and only results for the current branch are shown at a time. Observing the participants' interactions with the system showed that this behavior was only a limitation at the first attempt, but once they had familiarized themselves with the functionality, there was a noticeable improvement in how fluently they used the system. Interestingly, feedback to the static scope tree representation was more positive for experts in the second round, where the trees were larger and showed the structure of multiple files. It seems that these experts had an easier time understanding the representation (Q24) and judge the source code's complexity (Q25) within several files. It was mentioned that code structure allows only for partial assessment of the source code's complexity, since for example a single line of code can perform complex computations, and a single block can contain one or a hundred lines of code. Additional code metrics would be necessary. While some participants did not look at the SST at all, others wanted to use it but felt a lack of integration with other features, as well as difficulties comparing multiple trees. Based on this, we believe that additional features like using the SST as a code navigation tool, highlighting structures on code edits, and visualizing structural differences, can increase the utility of such a representation. Alternatively, more direct visualizations of the source code structure, e.g., in the form of a pixel-based visualization, should be taken into account.

The generally good responsiveness of the system when switching between states and investigating the different visual outputs and source code changes was positively mentioned by several participants. The feedback further improved in the second round, where caching of states led to even faster response times. The intermediate SUS score went up from 78.75, which corresponds to an adjective rating between "Good" and Excellent"~\cite{bangor2009determining}, to 88.75 ("Excellent"). This confirms our impressions and is a promising starting point for seeing our application in daily action. Quotes like "The programming task becomes more explorative and free." and "It is actually a lot of fun to go through the changes." emphasize our framework's investigation capabilities with respect to the algorithm's evolution. They show how our tool might positively influence the development workflow of visualization developers in the future.

The most sought after features were the comparison of states among different branches and the availability of the difference image for two specific states instead of showing it for collapsed groups only. Participants asked for a better link between the source code and the tooltip displaying code difference. This includes clicking on lines in the tooltip for navigation purposes, as well as highlighting code differences in the editor itself. Other feature requests included more control over how the states are grouped within the views, a merge tool, which combines the source code of two different states into one, and a way of deleting revisions. Otherwise, additional performance measures, more tools commonly seen in professional IDEs, and general customizability were requested.

\section{Discussion}
\label{sec:discussion}

Our proposed approach for designing a meta visualization system to provide insight into the evolution of the prototyping process of a scientific visualization algorithm
combines several interesting research fields. By utilizing methods from software visualization, visualization of time-oriented data, visualization of visualizations, and the analysis of the visual parameter space, it explores and raises new research questions. Based on the feedback we received, we believe it is a promising concept to provide visualization developers with novel information about their own research projects and a tool to investigate their own algorithm development. It further allows for an easier comparison of individual features to replace, improve, or combine them and enhance the visualization algorithm.  \\

\noindent\textbf{Ease of Use}

\noindent
By automatically compiling source code and visualizing the result in a live view, we follow in the footsteps of Vega-Lite~\cite{Satyanarayan2016} and ShaderToy~\cite{ShaderToy}. We not only apply the concept of an automatically updating result view to a programming language which requires compilation in the first place, but further visualize all previous results to emphasize on the result evolution. The fact that all intermediate steps are stored automatically further integrates with the idea of focusing the user's attention on the programming task.

The comparison of visualizations is a complicated task which has not yet been solved in a general manner. We facilitate the comparison of visualizations on a large and small scale by providing a juxtaposition and explicit encoding of all visual results and displaying their correlation to the underlying source code. We thereby increase the awareness of how visualizations evolve and in which way different features apply visual changes to the result. Our novel approach of applying the same parameter set to all visual results of the algorithm's states, further improves the comparability and allows for the visual exploration of the parameter space.\\

\noindent
\textbf{Scalability}

\noindent
While our system works well for developing prototypes for scientific visualization algorithms, it benefits from certain conditions given in this specific scenario. The visual result is commonly aligned to the original spatial dimensions in the data, which is often achieved via a virtual camera model projecting the spatial data into two-dimensional space. This alignment results in better comparability of different states based on images than in abstract visualizations, where vastly differing mappings from data to the visual result exist. Although the evaluation of our system has shown the experts' interest in utilizing our approach in other fields like information visualization or web development, it is unclear how the comparison task scales to these scenarios in a general manner. Additional study of this subject is required to extend our approach to other subfields in visualization.

Utilizing Diderot as a domain-specific language allows for comparatively short algorithms, that can be handled within a single source code file. When handling C++ and GLSL code, the structure and source code changes of multiple files need to be visualized, which results in both larger SSTs and longer tooltips. Scrolling along the time axis and tooltip does not scale indefinitely and if SSTs become very large, it is harder to make detailed comparisons between them. For example, SSTs in our use cases had a maximum depth of 5 and 6, and a maximum number of nodes of 7 and 51 respectively. In comparison, ParaView~\cite{ahrens2005paraview} as a full-fledged visualization application has a SST depth of 17 and ~40000 nodes in its core alone. An additional overview visualization, or a different visualization approach in general, would be necessary to support large multi-file applications with a lot of source code and many structural code changes.

We can reduce possible delays in the visual feedback by utilizing parallel execution capabilities of programming languages and compiling and executing several different states of the visualization algorithm at the same time on the server side. Looking at large scale development projects, compilation time and runtime become increasingly limiting factors for the visual support given. Since our system is built around short-term visual feedback and runs several revisions of the source code with possibly several parameter sets, the outlined benefits decrease for compilation- and runtime-heavy visual applications. While the revision tree, SST, visual result and source code difference are still available, the live view and parameter changes on older revisions would be delayed or might not be created at all within the time frame given between revisions.\\

\noindent
\textbf{Future Work}

\noindent
While creating a visual result from each compilable source code state includes all the interesting cases that create visual changes, it might produce many unnecessary results, take up a lot of screen space, and provide only little information. We try to counteract this issue by bundling results to compress the time axis of our meta visualization. However, at present we do not provide any interaction technique to remove revisions from the view, highlight interesting ones, or bundle states based on the user's interests. Solutions for the given tasks would be necessary when using our application for an extended period of time and when many revisions are created. It would be interesting to explore which other measures than the SST similarity could be utilized as factors for revision bundling and how such measures comply with the users' intent. Possible measurements might be computations of visual differences, number of code changes, or performance differences. Additional user interactions for tagging, removing or grouping states of interest might provide a good alternative to automatic approaches.

In the same manner, the difference image provides automatically generated information for improving the task of comparing several visualizations to each other. The exploration of other measures than image variance that are able to quantify other aspects of image difference, or improve the comparison task in different manners, might be a fruitful research area that we want to continue to investigate. We will further concentrate on the question of how such similarity measures can be correlated to source code to localize code constructs and their impact on the visual result.

While certain issues of an implementation might only become apparent for particular parameter sets, others can depend on the input data. Methods to compare and investigate the behavior of several algorithms on multiple input data sets need to be studied. Additionally, user tasks vastly vary, so that different supportive information about the source code and visualization need to be provided. Such options for task-specific customization for visualization experts need to be analyzed and integrated into our system. Since all the important information describing the visualization algorithm's evolution is stored on a server, many more interesting applications like building and sharing visual notebooks, or collaborating with multiple users on the same visualization algorithm come to mind. Our web client will be publicly available in such a way as to enable programmers to easily experiment with our interactive environment.\\

\noindent\textbf{Lessons Learned}

\noindent
Our system gives a direction for future development of visualization algorithms. Working with the system and evaluating it with experts in the field revealed several points of interest for future implementations of similar tools.

The accessibility of the system as a stand-alone website is particularly useful for teaching purposes, since no setup of applications and no prior knowledge is required to get the system running. The opposite is true for experts, who often have a running toolchain in place, have projects stored in existing repositories, and are used to their programming environment. They would greatly benefit from the visual support system being independent of the code editor and storage solutions. One could think of our system getting access to an existing repository and visualizing all available revisions, either as a web-based tool, or as a plugin to existing IDEs.

The evaluation made clear that different users have varying opinions on feature behavior. For the revision tree, some users preferred having the latest state at the top, instead of the bottom of the visualization. Users did not agree on if the source code difference should be shown from the hovered to the current or to the previous state. It could further be displayed vice versa, from the previous to the hovered state, which might depend on the task given. Some participants mentioned that the system was trying to update too often, although they were not finished editing the source code yet. These examples show that the system is required to provide extensive customization options to adjust the interface and feature behavior to the user's preferences.

The decision of using a modularized interface turned out very useful in terms of options for customization and extensibility. Following a similar approach on the server side allows for easy extension to support multiple programming languages. Overall, while our current prototype received positive user feedback and the proposed visualization and interaction methods were appreciated, we fully acknowledge that our system currently lacks several usability and customization features commonly found in professional IDEs. However, we plan to continue the development of our approach, with a specific focus on addressing its applicability to larger scale software projects.

\section{Conclusion}
\label{sec:conclusion}

We presented a novel approach for designing a meta visualization that enables the comparison of visual results of scientific visualization algorithms and their underlying source code at the same time. This concept yields additional insight into their relationship and thus enables programmers in the field of visualization to find correlations between visual changes and differences in the algorithms themselves.
Our approach supports programmers during the prototyping phase in keeping track of their development, while relieving them from repetitive tasks and thereby increasing their productivity. The frequent task of switching between different development states and comparing their visual outcome has been simplified to a one-click action by providing direct user interactions in the meta visualization. The problem of finding a given state with specific features is further supported through state identification via result images. We also showed how external functionality can be linked into the developed visualization algorithm, by providing an interactive view of the visual result and the on-the-fly coupling of specialized interaction functionality with program parameters. These parameters are applied to all states of the development process to enable the instant assessment of their impact. 